\documentclass{aa}  

\usepackage{graphicx}
\usepackage{caption}
\usepackage{subcaption}
\usepackage{natbib}
\bibpunct{(}{)}{;}{a}{}{,}
\usepackage{subcaption}
\usepackage{multirow}
\usepackage{xcolor}

\newcommand{\cii}{\mbox{[\ion{C}{ii}]}}
\newcommand{\oiii}{\mbox{[\ion{O}{iii}]}}

\newcommand{\lya}{Ly$\alpha$}

\usepackage{txfonts}

\begin{document} 

   \title{A tentative $\sim$1000 km s$^{-1}$ offset between the [CII] 158 $\mu$m and Ly$\alpha$ line emission  in a star-forming galaxy at $z = 7.2$}
   \titlerunning{Tentative [CII] 158 $\mu$m detection of GN-108036}
   
   \author{R. Baier-Soto,
          \inst{1}
          R. Herrera-Camus\inst{1,2},
         N. M. F{\"o}rster Schreiber\inst{3},
         A. Contursi\inst{4},
         R. Genzel\inst{3},
         D. Lutz\inst{3}, and
         L. Tacconi\inst{3}
          }

   \institute{Departamento de Astronom\'ia, Universidad de Concepci\'on, Barrio 			Universitario, Concepci\'on, Chile
              \label{1}
              \and
              Millennium Nucleus TITANS \label{2}
              \and
          Max-Planck-Institut f\"ur extraterrestische Physik (MPE), Giessenbachstr., D-85748 Garching, Germany	\label{3}
           \and
         Institute for Radio Astronomy in the Millimeter Range (IRAM), Rue de la Piscine, Grenoble, France	\label{4}
          }
          
 \date{\today}
 \authorrunning{Baier-Soto et al.}

 
  \abstract
   {GN-108036 is a star-forming galaxy at $z=7.21$, and one of the most distant known sources in the Northern hemisphere. Based on observations from the NOrthern Extended Millimeter Array (NOEMA), here we report the tentative detection of the \cii\ line at $\approx4\sigma$ significance. The integrated \cii\ line emission is spatially offset about $\sim4$ kpc from the rest-frame ultraviolet (UV) emission. The total \cii\ luminosity ($L_{\rm [CII]}=2.7\times10^8~L_{\odot}$) is consistent with the relation between \cii\ luminosity and star formation rate (SFR) observed in nearby and high-$z$ star forming galaxies. More interestingly, the \cii\ line is blueshifted with respect to the \lya\ line by $980\pm10$ km s$^{-1}$. If confirmed, this corresponds to the largest velocity offset reported to date between the \lya\ line and a non-resonant line at $z\gtrsim6$. According to trends observed in other high redshift galaxies, the large \lya\ velocity offset in GN-108036 is consistent with its low \lya\ equivalent width and high UV absolute magnitude. Based on \lya\ radiative transfer models of expanding shells, the large \lya\ velocity offset in GN-108036 could be interpreted as the presence of a large column density of hydrogen gas, and/or an outflow with a velocity of $v_{\rm out}\sim\Delta v_{\rm Ly \alpha}/2\sim500$ km s$^{-1}$. We also report the 3$\sigma$ detection of a potential galaxy companion located $\sim30$ kpc east of GN-108036, at a similar systemic velocity, and with no counterpart rest-frame UV emission. 
   }
   
   

   \keywords{Galaxies: high-redshift - Galaxies: ISM - Galaxies: star formation}

   \maketitle

\section{Introduction}

The first galaxies most likely formed during the first $\sim$~200-300~Myr of the Universe lifetime \citep[e.g.,][]{Bromm2011, wise2011birth}. These early galaxies represent the primordial building blocks of the galaxy population we observe today. During that early epoch ($z\gtrsim8$), the gas in the Universe was mostly neutral, which makes the first galaxies, and their increasing star formation activity, natural contributors to the reionization of the Universe \citep[e.g.,][]{fan2006constraining}. To better understand the properties of these first systems, it is important to consider a multi-wavelength approach, that takes into account the interplay between stars, dust, and warm and cold gas.




Until recently, the study of the first galaxies was limited to the characterization of its nebular and stellar emission: \textit{Hubble Space Telescope} (HST) near-infrared observations of young and massive stars, \textit{Spitzer} mid-infrared observations of the bulk of the stellar population, and rest-frame ultraviolet (UV) observations from ground telescopes of the \lya\ and higher ionization lines 
\citep[e.g.,][]{ono2011spectroscopic, Zitrin2015,Stark2017}. Over the last decade, and thanks to the advent of the improved capabilities of the NOrthern Extended Millimeter Array (NOEMA) and the Atacama Large Millimeter/sub-millimeter Array (ALMA), we now have access to the study of the cold and neutral gas component in these early systems.  



\begin{figure*}[ht!]
\centering
\begin{subfigure}{.5\textwidth}
  \centering
  \includegraphics[scale=0.30]{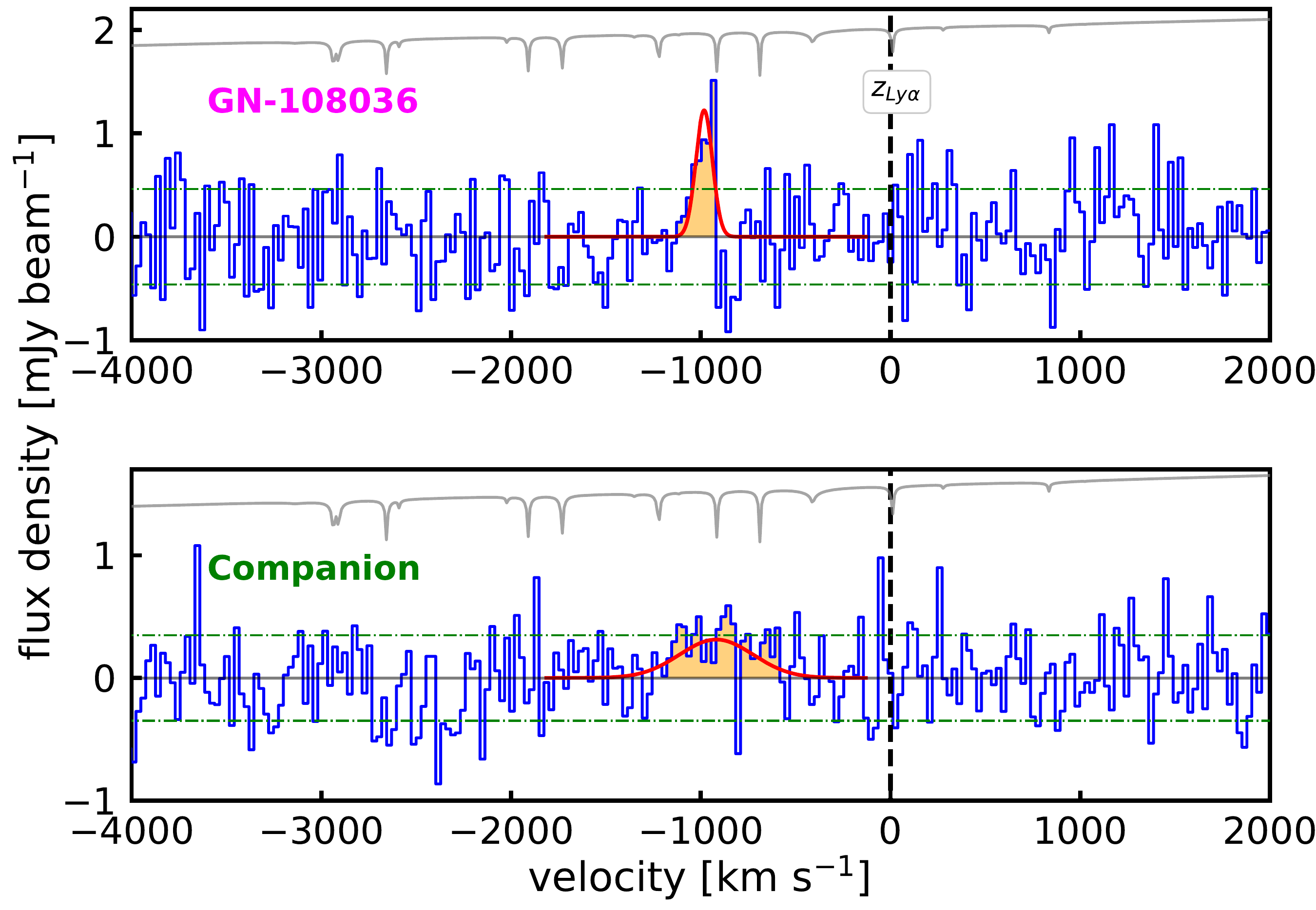}
\end{subfigure}%
\begin{subfigure}{.5\textwidth}
  \centering
  \includegraphics[scale=0.33]{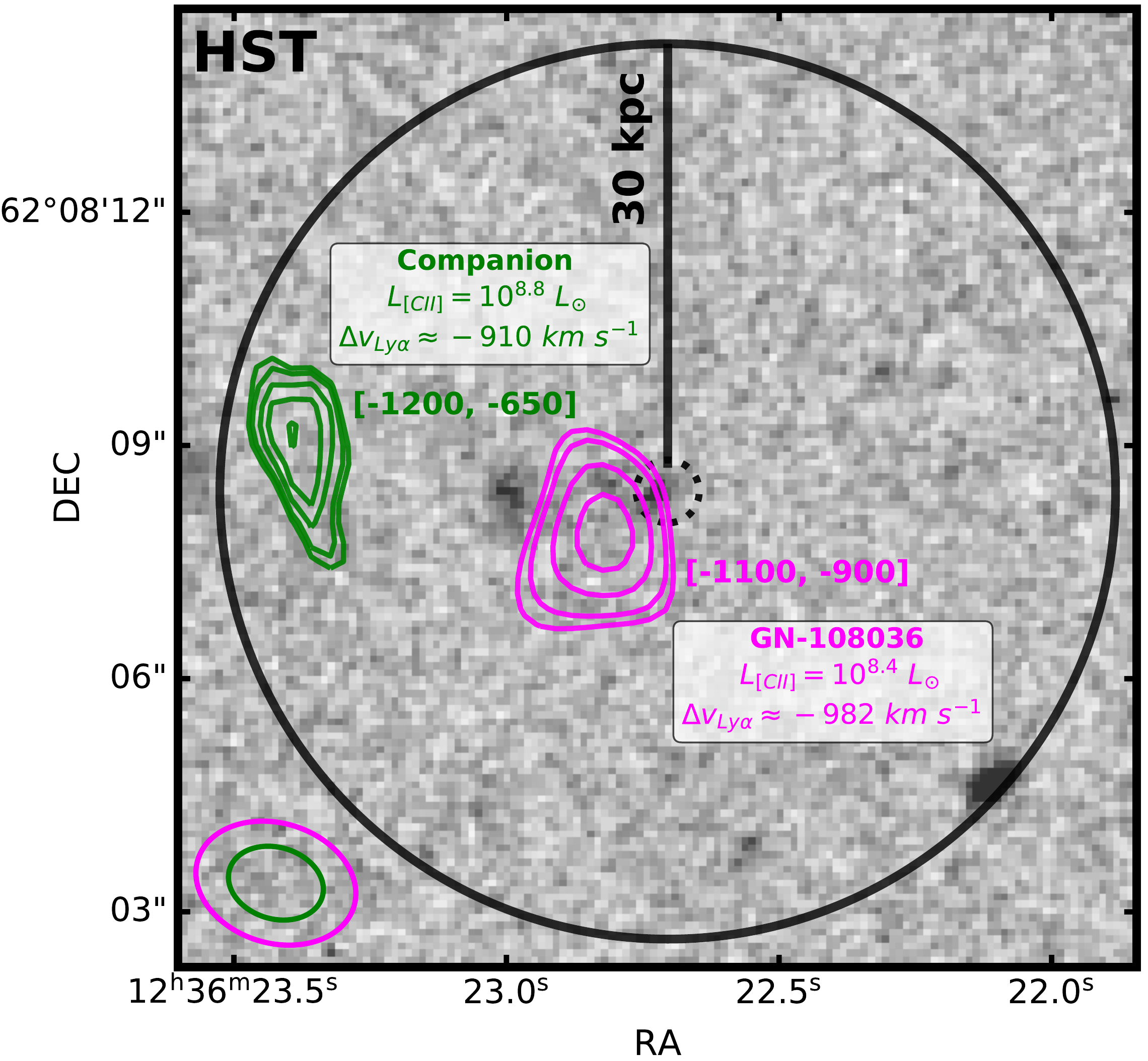}
\end{subfigure}
\caption{\textit{Left:} \cii\ line spectrum of GN-108036 (top) and a potential companion located $\sim30$ kpc east (bottom). The zero velocity is set using the redshift from the detection of the Ly$\alpha$ line \citep[black dashed vertical line][]{ono2011spectroscopic}. The best 1D Gaussian fits to the tentative detections are shown in red. The spectrum of GN-108036 and the potential companion are extracted from the compact and extended array configuration data, respectively. \textit{Right:} HST rest-frame UV map of the field of GN-108036. The black dotted circle indicates the position of the HST rest-frame UV emission from GN-108036, and the black solid circle around the center has a radius of 30 kpc. The integrated intensity contours of the \cii\ line emission in GN-108036 (2.5, 3, 4 and 5$\sigma$ levels) extracted from the compact array data are shown in magenta. In green we show the potential detection of a companion based on the \cii\ line emission extracted from the extended array data. The beam sizes are shown in the lower left corner.  The atmospheric transmission curve in the frequency range covered by our observations is shown as a gray line on the top part of the panels.\label{fig:fit_contour}}
\end{figure*} 

The main tracer of the cold gas in high-$z$ galaxies is the \cii\ 157.74 $\mu$m fine structure transition, one of the major coolants of the neutral gas \cite[e.g.,][]{wolfire03}. One advantage of the \cii\ line is that is bright (typically $\sim$ $0.1$-1\% of the far-infrared luminosity \citep[e.g.,][]{stacey1991,herrera2018shining}, and remains bright in metal-poor environments \citep[e.g.,][]{israel2011,cormier2014,cigan2015,bouwens2021reionization}. Because the C ions can be collisionally excited by hydrogen atoms and molecules, the \cii\ line represents a powerful alternative to trace the neutral gas. This is particularly relevant at high-$z$ given the difficulties or serious limitations to observe the CO and HI transitions. At $z\gtrsim6$, \cii\ line observations of star-forming galaxies have revealed clumpy gas structure, which is typically spatially offset from the UV emission, and follows in general the observed relation between the star formation rate (SFR) and the \cii\ luminosity observed in nearby galaxies \citep[e.g.,][]{maiolino2015,carniani2017,2018ApJ...854L...7C}. 



In combination with the \cii\ transition, another important tracer of these very high-$z$ systems is the \lya\ line, produced by young massive stars. The \lya\ line is resonant, therefore, it is typically offset in velocity with respect to non-resonant lines \cite[e.g.,][]{steidel10,2014ApJ...795...33E,hashimoto15,2020A&A...643A...6C}. The observed \lya\ line structure offers valuable information about the interstellar medium (ISM) and it surrounding intergalactic medium (IGM). For example, blueshifted or redshifted \lya\ emission with respect to the systemic redshift of a system could indicate the presence of inflowing or outflowing gas, respectively \citep[e.g.,][]{dijkstra06,verhamme06,gronke15}. A compilation by \cite{hashimoto2019big} of \lya, \cii, and \oiii\ 88 $\mu$m line observations of $z\approx6-8$ star-forming galaxies, shows that these systems tend to have \lya\ velocity offsets in the $\approx100-500$ km s$^{-1}$ range, and that galaxies with the largest velocity offsets have lower \lya\ equivalent widths and higher star formation rates. Based on simple expanding spherical shell models, these large velocity offsets could be interpreted as if the galaxy has a large neutral hydrogen column density and/or an outflow \citep[e.g.,][]{verhamme06,verhamme15}.

In this letter we focus on the combined analysis of the \lya\ and \cii\ line emission GN-108036, a star-forming galaxy at $z=7.213$. The redshift of GN-108036 is based on the detection of the \lya\ line from deep Keck/DEIMOS spectroscopy \citep{ono2011spectroscopic}. This makes GN-108036 one of the most distant sources known in the Northern Hemisphere (dec. +62:08:07). The stellar mass of GN-108036 is $M_{\star}=10^{8.76}$~$M_{\odot}$, and its star formation rate (SFR) is in the 29 to 100~$M_{\odot}~{\rm yr}^{-1}$ range based on calculations using the rest-frame UV continuum and stellar population synthesis models, respectively \citep{ono2011spectroscopic}. 


This work is organized as follows. In Section 2 we describe the observations and data reduction. In Section 3 we present the results.  In Section 4 we discuss the \cii\ properties of the galaxy and the velocity offset with respect to the \lya\ line emission. In Section 5 we present the summary and conclusions. For this work we adopt the following cosmological parameters: $H_0=67.4 \ \mbox{km s}^{-1} \ \mbox{Mpc}^{-1}$, $\Omega_{M}=0.315$ and $\Omega_{\Lambda}=0.685$ \citep{2020A&A...641A...6P}. For a source at $z=7.21$, this results in a physical scale of 5.24 kpc/$''$.

\section{Observations and data reduction}

We used NOEMA to observe GN-108036 in the \cii~158 $\mu$m transition and dust continuum. At the redshift of the source, the \cii\ transition is redshifted to $\nu_{\rm [CII],obs}=231.5$~GHz, which falls into NOEMA Band 3. GN-108036 was first observed in March 2019 using the most compact array configuration (D) for an on-source time of 3.2~hrs. The second set of observations was taken on March 2020 using array configuration C for an on-source time of 3.7~hrs. We reduced and combined both data sets using the \texttt{CLIC} and \texttt{MAPPING} software by IRAM.\footnote{CLIC and MAPPING are part of the GILDAS package \citep{Guilloteau00}: \url{http://www.iram.fr/IRAMFR/GILDAS}}. For the imaging of the \cii\ cube and the dust continuum map we use natural weighting to maximize the sensitivity. The resulting synthesized beam for the D, C, and combined C+D data was $\theta=2.1'' \times1.5''$, $\theta=1.2''\times0.9''$, and $\theta=1.4''\times1.1''$, respectively. The rms noise for the D, C, and combined C+D line cubes is 0.46, 0.35, and 0.35 mJy beam$^{-1}$ in 25 km s$^{-1}$ channels, respectively.

We also created a dust continuum map using part of the sidebands of the C+D data where we do not expect line emission from the source. The rms noise in this map is 13 $\mu$Jy beam$^{-1}$. Assuming a characteristic dust temperature for a $z\sim6-7$ galaxy of $T_{\rm dust}=45$ K \citep[e.g.,][]{schreiber2018,faisst2020}, and a dust emissivity index of $\beta=1.5$, the expected non detection indicates a dust mass upper limit of $M_{{\small\mbox{dust}}}<9.5 \times 10^6$ $M_{\odot}$. The low dust content in GN-108036 is consistent with that observed in  other massive ($M_{\star}\sim 10^{9} M_\odot$), star-forming galaxies at $z\sim7-9$, including: A2744-YD4 at $z= 8.4$ \citep[$M_{\rm dust}\approx6\times10^{6} M_\odot$; ][]{laporte17}, B14-65666 at $z= 7.2$ \citep[$M_{\rm dust}\approx10^{7} M_\odot$; ][]{hashimoto2019big}, A1689-zD1 at $z= 7.13$ \citep[$M_{\rm dust}\approx2\times10^{7} M_\odot$; ][]{Bakx2021}, and a handful of luminous Lyman-break galaxies at $z\sim7-8$ \citep[$M_{\rm dust}\lesssim5\times10^{7} M_\odot$ if the dust temperature is $\gtrsim40$ K;  ][]{Schouws2022}.



\section{Results}

\subsection{Tentative detection of the \cii\ 158~$\mu$m transition in GN-108036}

We performed a blind search for \cii\ line emission by systematically placing apertures of the beam size across the cubes separated by a distance of a quarter of a beam size. We tentatively detected two sources with an integrated signal-to-noise (S/N) of $\gtrsim3$ in two regions of the cube: 1) in the center, and slightly offset from the spatial position of the detection of GN-108036 in the HST rest-frame UV and Ly$\alpha$ data, and 2) about $\sim$30 kpc east from the HST detection of GN-108036. 

The top-left panel of Fig.~\ref{fig:fit_contour} shows the \cii\ spectrum of the potential detection of GN-108036 extracted from the compact array NOEMA data. From a single Gaussian fit we find that the line is centered at $-982\pm13$ km s$^{-1}$ with respect to the detection of the Ly$\alpha$ line \citep{ono2011spectroscopic}. We discuss more about this large velocity offset in Section \ref{sec:ly_offset}. The curve of atmospheric transmission overploted as a gray line shows that the tentative line detection is not a result of a strong or broad atmospheric absorption line. The integrated \cii\ flux is 0.22$\pm$0.06 Jy km s$^{-1}$, which corresponds to a detection of the source with a S/N of $3.7$.  In Appendix \ref{appB} we also show the histogram of the peak S/N per beam in the compact array data. As expected, the distribution roughly follows a Gaussian shape, and the potential detection of GN-108036 with a peak S/N of 5.4 (magenta bin) corresponds to a high-S/N ``outlier''.

The left panel of Fig. \ref{fig:spectras} in the Appendix shows the spectra extracted in the same region from the extended array and combined array data. The signal is present in the $\sim2\times$ higher angular resolution dataset at the same velocity range, but with lower significance. This could be the result of the \cii\ line emission in GN-108036 to be significantly more extended than $1\arcsec$ ($\sim5$~kpc), as it has been observed in other $z\sim6-7$ star-forming galaxies \citep[e.g.,][]{carniani2020}. 


\begin{table*}[htbp!]
\caption{\cii\ 158 $\mu$m fluxes and parameters from the Gaussian fit to the tentative detections of GN-108036 and its companion}             
\label{table:2}      
\centering          
\begin{tabular}{c c c c c c c c}
\hline \hline
Source & SFR & $z_{\rm Ly\alpha}$ & $z_{\rm [CII]}$ & Central velocity  & FWHM  & Integrated Flux & Luminosity \\ 
& [$M_{\odot}$ yr$^{-1}$] & & & [km s$^{-1}$] & [km s$^{-1}$] & [Jy km s$^{-1}$] & 10$^{8} L_{\odot}$ \\
\hline   
   GN-108036 & $\sim30-100$ & 7.213 & 7.180 & $-982.2$ $\pm$ 12.7   & 102.7 $\pm$ 29.9 &  0.22 $\pm$ 0.06 & 2.7 \\
 Companion & $-$ & $-$ & 7.188 & $-910.1$ $\pm$ 57.2 & 503.5 $\pm$ 134.7 & 0.47 $\pm$ 0.15 & 6.0 \\
\hline
\end{tabular}
\label{tab:gaussian_parameters}
\end{table*}

We constructed a \cii\ integrated intensity map (or moment 0) integrating the \cii\ line emission around the potential detection of the line centered at the velocity of $-982$ km s$^{-1}$.The peak signal-to-noise in the integrated \cii\ line emission map is \textbf{$\approx5.4$}
. Fig.~\ref{fig:flux_maps} in the Appendix shows the \cii\ moment 0 map, and the right panel of Fig.~\ref{fig:fit_contour} shows the \cii\ integrated intensity contours (at significance levels of 2.5, 3, 4 and 5$\sigma$) overplotted on the HST/WFC3 map of the field. The dotted black circle in the center indicates the position of  GN-108036 as detected in the rest-frame UV and Ly$\alpha$ emission \citep[][]{ono2011spectroscopic}. The peak of the integrated \cii\ line emission is offset with respect to the peak of the rest-frame UV and Ly$\alpha$ emission by $\sim4$ kpc in the south-east direction. We checked the astrometric accuracy of the HST images using stars in the field in the GAIA catalog \citep[][]{gaia_2018}, and this is not the source of the observed offset. Spatial offsets between the star-forming regions and \cii\ line emission have been observed in other star-forming galaxies at $z\gtrsim5$ \citep[e.g.,][]{2018ApJ...854L...7C}, and could be related to difference in the ionizing state of the gas, dust obscuration, and/or the effect of stellar feedback destroying molecular gas \citep[e.g.,][]{vallini15,katz17}. 

Together with the tentative detection of the  
\cii\ line in GN-108036, we identify a potential additional system located  approximately at 30 kpc in the east direction. The spectrum is shown in the lower-left panel of Fig.~\ref{fig:fit_contour}. Interestingly, the tentative detection is at a similar velocity ($-910\pm57$ km s$^{-1}$) of the possible detection of the \cii\ line in GN-108036, but the line profile is significantly wider ($503\pm134$ km s$^{-1}$). The integrated \cii\ flux is 0.47$\pm$0.15 Jy km s$^{-1}$, which corresponds to a tentative detection with a S/N of $\approx3$. As Fig. \ref{fig:flux_maps} in the Appendix shows, the signal is much weaker in the compact array data. The contours of the integrated \cii\ line emission from the extended array data are shown in green in the right panel of Fig.~\ref{fig:fit_contour}. 


Table \ref{tab:gaussian_parameters} summarizes the \cii\ line properties of the tentative detections of GN-108036 and the companion. We include the redshift of the source (Ly$\alpha$ and \cii), the central velocity and FWHM of the \cii\ line from the best 1-D Gaussian fit, the integrated \cii\ flux, and the \cii\ luminosity.

\begin{figure}[t!]
    \centering
    \includegraphics[width=0.9\linewidth]{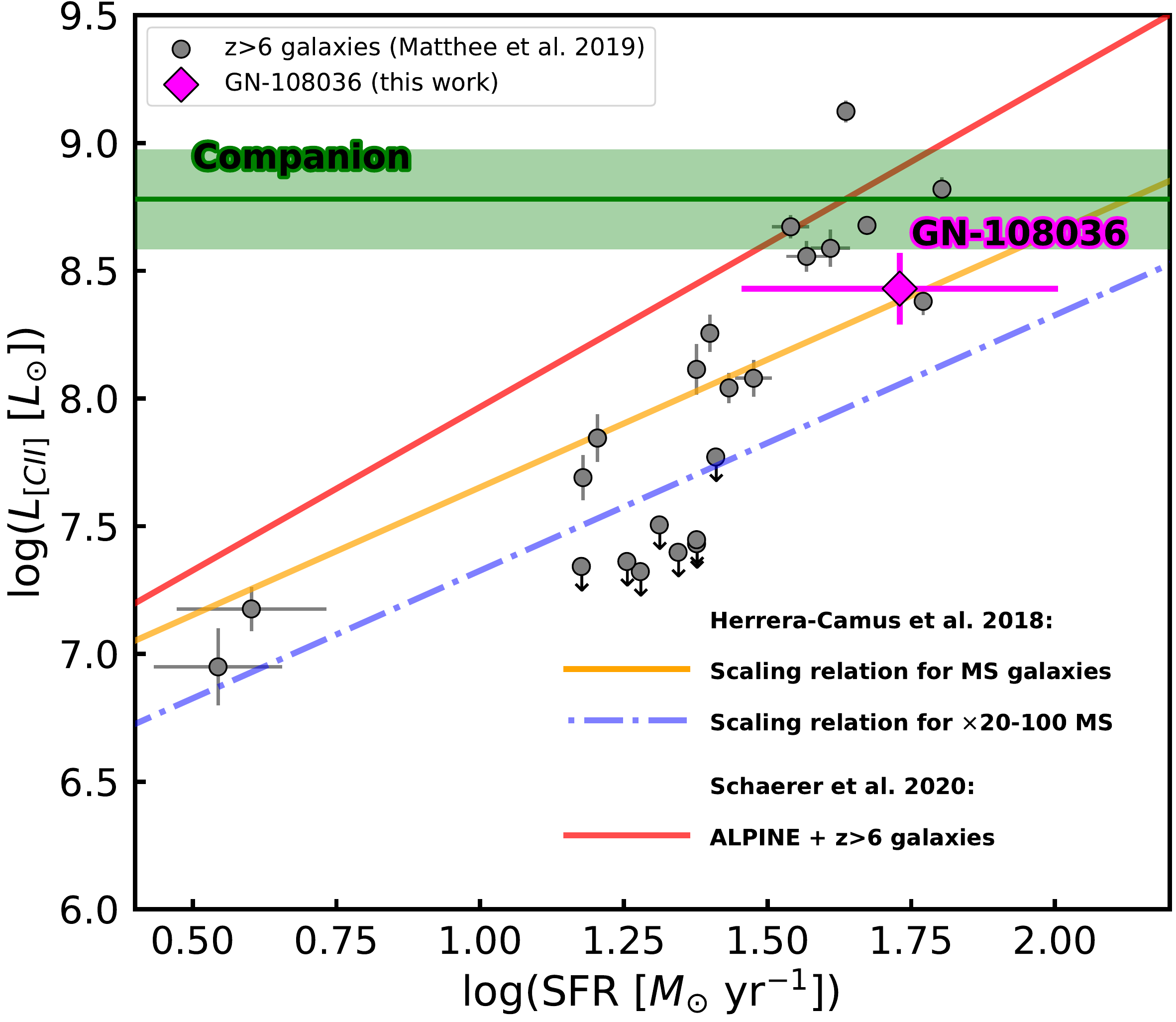}
    \caption{\cii\ luminosity $-$ SFR relation observed in $z\gtrsim6$ galaxies \citep[gray circles;][]{Matthee2019}. The solid (orange) and dot-dashed (blue) lines correspond to the scaling relations observed in star-forming galaxies on and above ($\times20-100$) the main-sequence independent of redshift \citep[][]{herrera2018shining}, and the red solid line shows the best-fitting relation based on ALPINE ($z\sim4-5$) and $z\gtrsim6$ galaxies \citep{schaerer20}. The tentative detection of GN-108036 is shown  with a magenta diamond. The SFR of GN-108036 ranges from $\approx30$ to $\approx100$ M$_{\odot}$~year$^{-1}$ depending if the rest-frame UV emission or SED are used, respectively \citep{ono2011spectroscopic}. The green solid line indicates the \cii\ luminosity for the companion, which does not have an HST counterpart or SFR estimate available.
    \label{fig:SFR_relation}}
\end{figure}

\section{Analysis}

\begin{figure*}[htbp]
    \centering
    \includegraphics[scale=0.4]{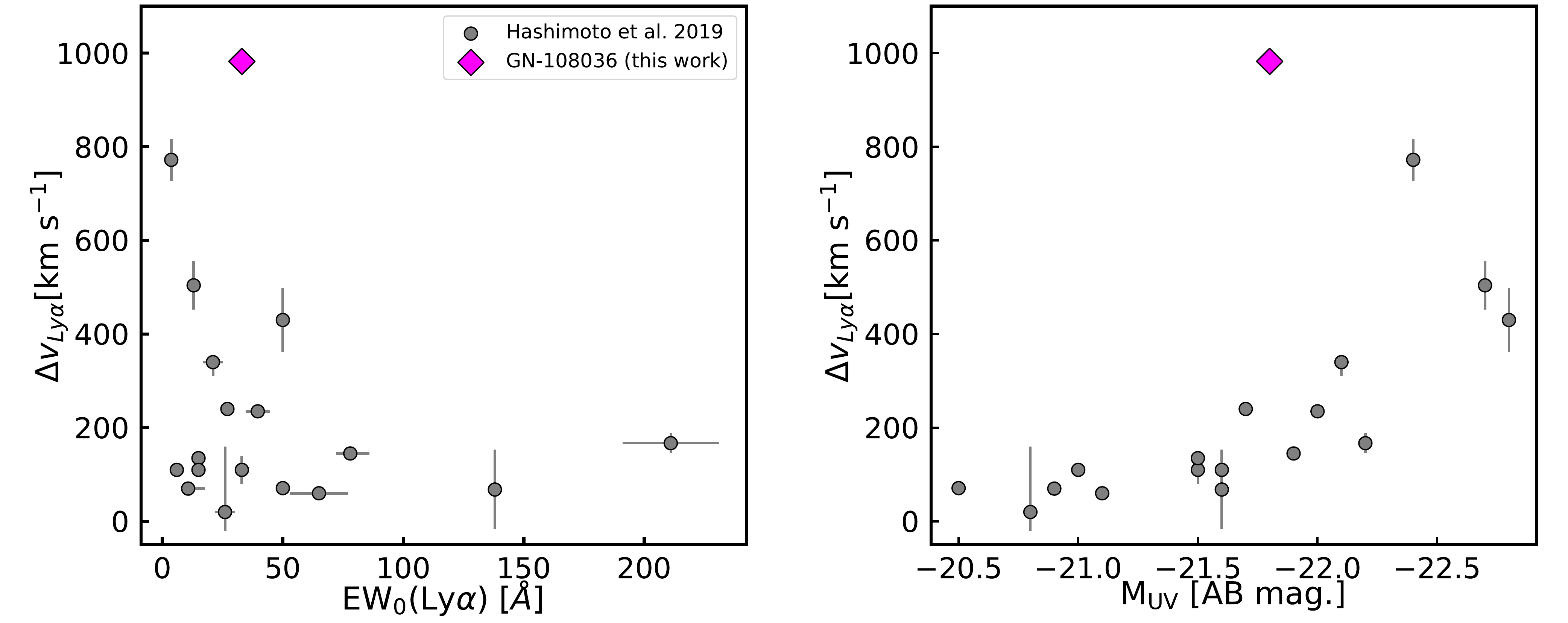}
    \caption{\lya\ velocity offset ($\Delta v_{Ly \alpha}$) with respect to the \cii\ line as a function of \lya\ equivalent width (\textit{left}) and absolute UV magnitude (\textit{right}) observed in star-forming galaxies at $5<z<8$ \citep[gray circles; ][]{hashimoto2019big}. The tentative detection ($3\sigma$) of GN-108036 is shown in both panels with a pink diamond.
    }
    \label{fig:lyoff}
\end{figure*}

\subsection{\rm Relation between \cii\ line emission and star formation activity} 
\label{sec:ly_offset}

Under the assumption of an interstellar medium in thermal equilibrium, and considering that the \cii\ transition is one of the main cooling channels \cite[e.g.,][]{wolfire03}, a tight relation is expected between the \cii\ line emission and the star formation activity. This relation has been observed in nearby, star-forming galaxies and high-$z$, main-sequence systems \cite[e.g.,][]{delooze14,rhc15,herrera2018shining,schaerer20}. Fig. \ref{fig:SFR_relation} shows the \cii\ luminosity - SFR relation for star-forming galaxies detected at $z\gtrsim6$ \citep{Matthee2019}. The solid (orange) and dot-dashed (blue) lines represent the best fit to star-forming galaxies on and above ($\times20-100$) the main-sequence \citep[corrected by the redshift dependence of the main-sequence;][]{herrera2018shining}, and the red line represents the best fit to ALPINE ($z\sim4-5$) and other $z\sim6-8$ galaxies \citep{schaerer20}. GN-108036, with SFR estimates from different indicators ranging between $\sim$30 and 100~$M_{\odot}~{\rm yr}^{-1}$ \citep{ono2011spectroscopic}, follows the main relation observed in other $z\gtrsim6$ galaxies, and lies in between the \cii$-{\rm SFR}$ scaling relations for galaxies on and above the main-sequence. Regarding the potential companion of GN-108036, there is no HST counterpart or SFR estimate available, so we include the \cii\ luminosity value as an horizontal green line.\footnote{ All measurements and scaling relations in Figure \ref{fig:SFR_relation} have been scaled to the same initial mass function  \citep{salpeter55} following the conversion factors listed in \cite{madau2014}. }

The fact that GN-108036 follows the \cii$-{\rm SFR}$ relation observed in other $z\gtrsim6$ star-forming galaxies, combined with the small spatial offset observed between the peak of the \cii\ line and the rest-frame UV and Ly$\alpha$ emission, argues in favor of the interpretation of the \cii\ line detection in GN-108036 as real and associated with the galaxy. 

\subsection{Ly$\alpha$ - \cii\ velocity offset} 
\label{sec:ly_offset}

\lya\ is a resonant line, thus its profile carries important information about the content, geometry and kinematics of the atomic gas. 
At $z\approx 2-3$, star-forming galaxies can show significant velocity differences between \lya\ and non-resonant lines (e.g., H$\alpha$, H$\beta$, \oiii) that range between 100 to 1000~km~s$^{-1}$ \citep[e.g., ][]{2013ApJ...775..140H,2014ApJ...795...33E}. At $z\gtrsim6$, Lyman Break galaxies show \lya\ velocity offsets with respect to the \cii\ line that are typically between 100 to 500~km~s$^{-1}$. The record belongs to the star-forming galaxy B14-65666 at $z=7.15$, with \lya\ line emisison redshifted with respect to the \cii\ and \oiii\ lines by $\Delta v_{\small \mbox{Ly}\alpha}=772$~km~s$^{-1}$ \citep{hashimoto2019big}.

In the case of GN-108036, the tentative detection of the \cii\ line is blueshifted with respect to the \lya\ line by $982.2\pm12.7$~km~s$^{-1}$, the largest velocity offset reported to date for a system at $z\gtrsim6$. Figure \ref{fig:lyoff} compares the \lya\ velocity offset in GN-108036 with star-forming galaxies at $z\gtrsim6$ compiled by \cite{hashimoto2019big}. The left panel shows the anti-correlation observed between $\Delta v_{\small \mbox{Ly}\alpha}$ and \lya\ equivalent width (EW$_0(\mbox{\lya})$), and the right panel shows the positive correlation observed between $\Delta v_{\small \mbox{Ly}\alpha}$ and the UV absolute magnitude ($M_{UV}$) of the system. 


To first order, and based on models of \lya\ radiative transfer in expanding shells, there are two scenarios that can explain the large velocity offset observed in GN-108036. In the first scenario, the presence of a large column density of atomic hydrogen implies that \lya\ photons suffer from more dust attenuation due to a larger optical path length, which causes a reduction of the \lya\ equivalent width and an increase in the \lya\ velocity offset \citep[e.g.,][]{2014ApJ...795...33E}. In the second scenario, the increasing UV absolute magnitude is correlated with stronger star formation activity, which can drive outflows including atomic gas that would increase the \lya\ velocity offset. In a simple approximation, it is expected that the velocity of the outflow ($v_{\mbox{{\small out}}}$) is correlated with $\Delta v_{\rm Ly \alpha}$ as $\Delta v_{\rm Ly \alpha} \sim 2 \times v_{\mbox{{\small out}}}$ \citep[e.g.,][]{verhamme06}. This would imply an atomic gas outflow velocity for GN-108036 of $\sim500$~km~s$^{-1}$. This outflow velocity is consistent with those observed in local starburst with comparable levels of star formation activity \cite[e.g.,][]{shapley2003,Heckman16}.





\section{Summary and conclusions}

We report new NOEMA Band 3 observations of the \cii\ 158 $\mu$m transition and dust continuum in one of the most distant sources in the Northern hemisphere, the star-forming galaxy GN-108036 detected in \lya\ emission at $z=7.12$ \citep{ono2011spectroscopic}. Our main results can be summarized as follows:

\begin{enumerate}

    \item We tentatively detect GN-108036 in \cii\ line emission with a $S/N\approx4$. The peak of the integrated emission is spatially offset about 4 kpc with respect to the peak of the rest-frame UV and \lya\ line detection \citep{ono2011spectroscopic}. Spatial offsets of similar magnitudes are commonly observed in star-forming systems at $z\gtrsim6$ \citep[e.g.,][]{2018ApJ...854L...7C}. The potential  \cii\ detection is blueshifted with respect to the \lya\ emission by $982.2\pm12.7$ km s$^{-1}$. If confirmed, this would be the largest Lya velocity offset reported to date for a $z\gtrsim6$ star-forming galaxy. GN-108036 is not detected in the dust continuum, and the $3\sigma$ dust mass upper limit is $M_{{\small\mbox{dust}}}<9.5 \times 10^6$ $M_{\odot}$. 
    
    \medskip
    
    \item Together with GN-108036, we tentatively detect ($3\sigma$) in \cii\ line emission one additional source at similar systemic velocity but located $\approx30$ kpc east of GN-108036. This source has no counterpart in the HST imaging of the field. 
    
    \medskip
    
    \item GN-108036, with a SFR that ranges between $\sim30-100$~$M_{\odot}~{\rm yr}^{-1}$ \citep{ono2011spectroscopic}, follows the relation between the \cii\ luminosity and the SFR observed in star-forming galaxies at $z\gtrsim6$ \citep[e.g.,][]{Matthee2019}, and is consistent with the scaling relations of $L_{\rm [CII]}-{\rm SFR}$ observed in nearby and high-z main-sequence star-forming galaxies  \citep[e.g.,][]{herrera2018shining,schaerer20}. The fact that the potential \cii\ emission in GN-108036 is almost co-spatial with the rest-frame UV and \lya\ emission, and that GN-108036 follows the $L_{\rm [CII]}-{\rm SFR}$ relation, argues in favor of the \cii\ line detection to be real.
    
    \medskip
    
    \item The \lya\ velocity offset observed in GN-108036 is consistent with the positive and negative correlations observed between $\Delta v_{Ly \alpha}$ and EW$_0(\mbox{\lya})$ and $M_{\rm UV}$ in $z\gtrsim6$ star-forming galaxies, respectively. If models of \lya\ radiative transfer in expanding shells apply to GN-108036, the physical scenarios that could explain the observed large Lya velocity offset, the low EW$_0(\mbox{\lya})$ and high $M_{\rm UV}$ are: 1) the presence of a large HI column density, 2) the existence of an outflow with velocity $v_{\rm out}\sim\Delta v_{\rm Ly \alpha}/2\sim500$ km s$^{-1}$. Certainly deeper, higher angular resolutions observations of GN-108036 are needed to confirm the \cii\ line detection, and further explore these two scenarios.
    
\end{enumerate}

The upgraded NOEMA capabilities, which will have 12 antennas by the end of the summer of 2022, and has a correlator ($PolyFiX$) with a bandwidth of $\sim31$ GHz, offers a great opportunity to search and detect in \cii\ line emission $z\gtrsim6$ galaxies based on robust photometric redshifts estimates. The latter should become available in large numbers in the near future thanks to the {\it James Webb Space Telescope}.

\begin{acknowledgements}
We thank the referee for very useful comments and suggestions that improved the manuscript. R. B.-S and R.H.-C. thank the Max Planck Society for support under the Partner Group project "The Baryon Cycle in Galaxies" between the Max Planck for Extraterrestrial Physics and the Universidad de Concepción. R.H-C also acknowledge financial support from Millenium Nucleus NCN19\_058 (TITANs) and support by the ANID BASAL projects ACE210002 and FB210003.
\end{acknowledgements}

\bibliographystyle{aa}
\bibliography{references}

\begin{appendix}

\section{Significance of the potential \cii\ detections of GN-108036 and its companion} \label{appB}

\begin{figure}[ht!]
\centering
  \includegraphics[scale=0.43]{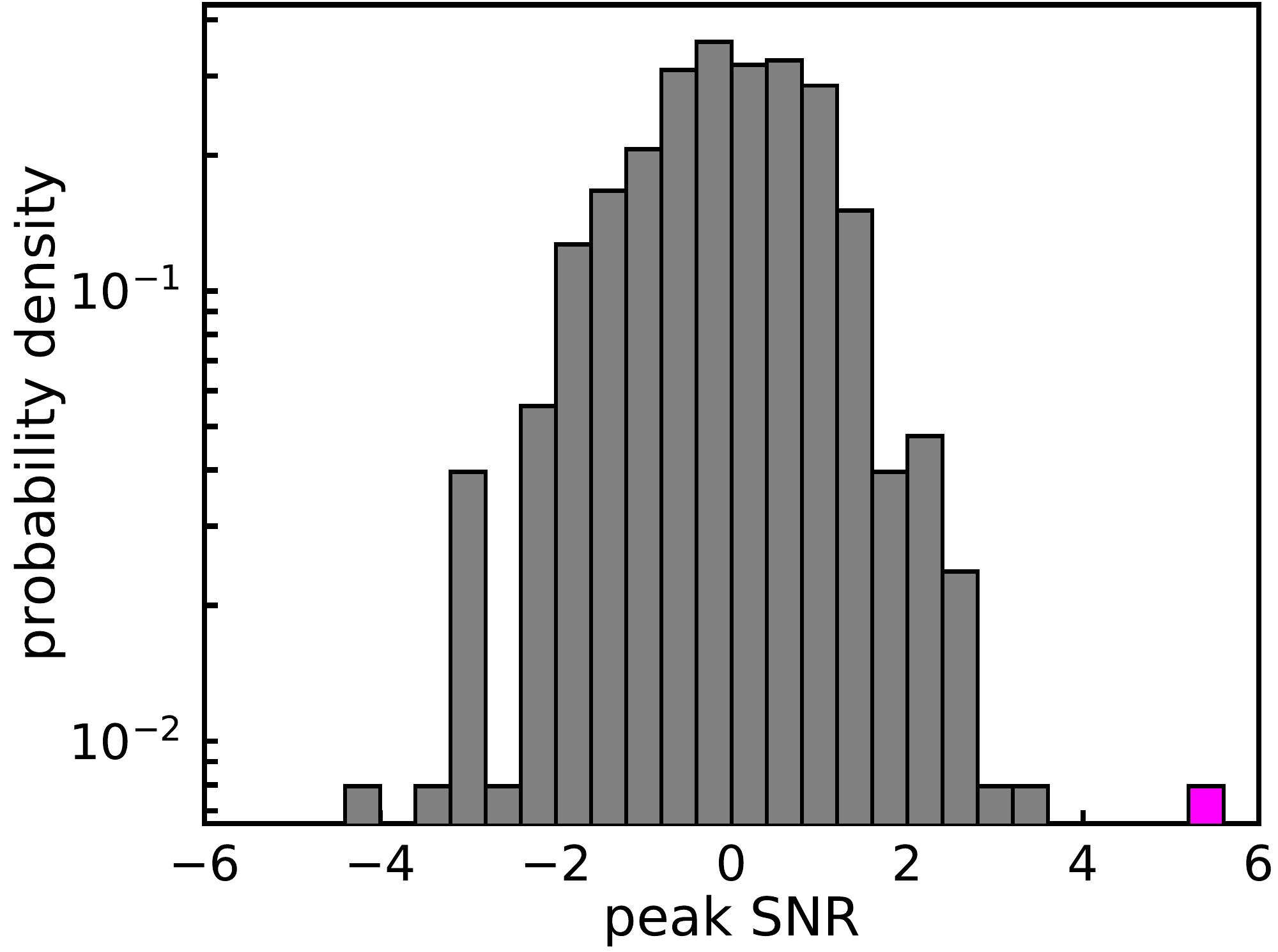}
\caption{\textit{Left:} Distribution of the peak S/N values (positive and negative) for the compact array data. The tentative detection of GN-108036 is shown as magenta bin.}
\label{fig:histograms}
\end{figure}

 Fig. \ref{fig:histograms} shows the  distribution of positive and negative peak S/N values per beam of the compact array data, respectively. The tentative detection of GN-108036 is shown as a magenta bin with a peak S/N value of 5.3.

\section{NOEMA \cii\ line observations of GN-108036 with different array configurations}

\begin{figure*}[ht!]
\centering
\begin{subfigure}{.5\textwidth}
  \centering
  \includegraphics[scale=0.3]{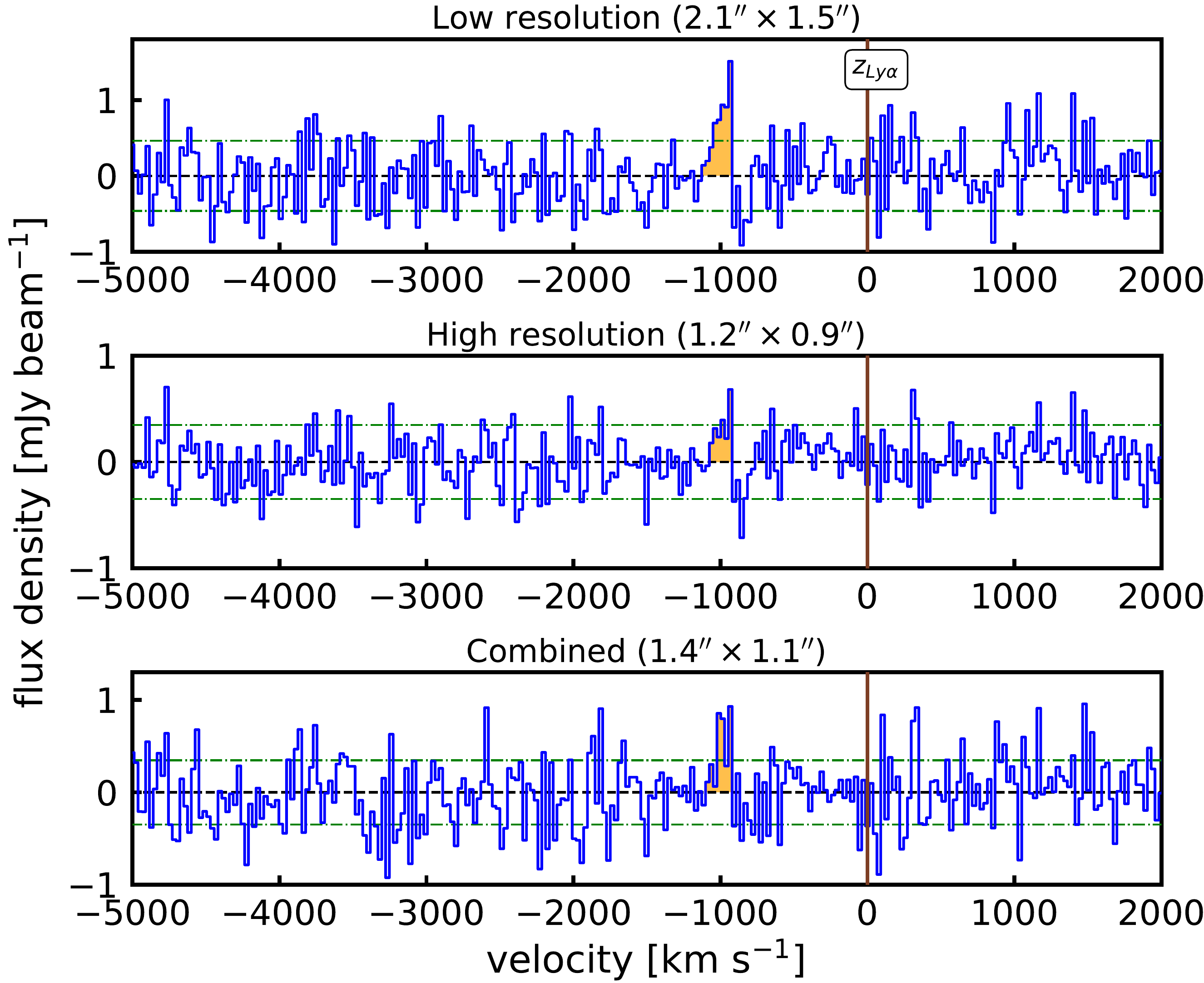}
\end{subfigure}%
\begin{subfigure}{.5\textwidth}
  \centering
  \includegraphics[scale=0.3]{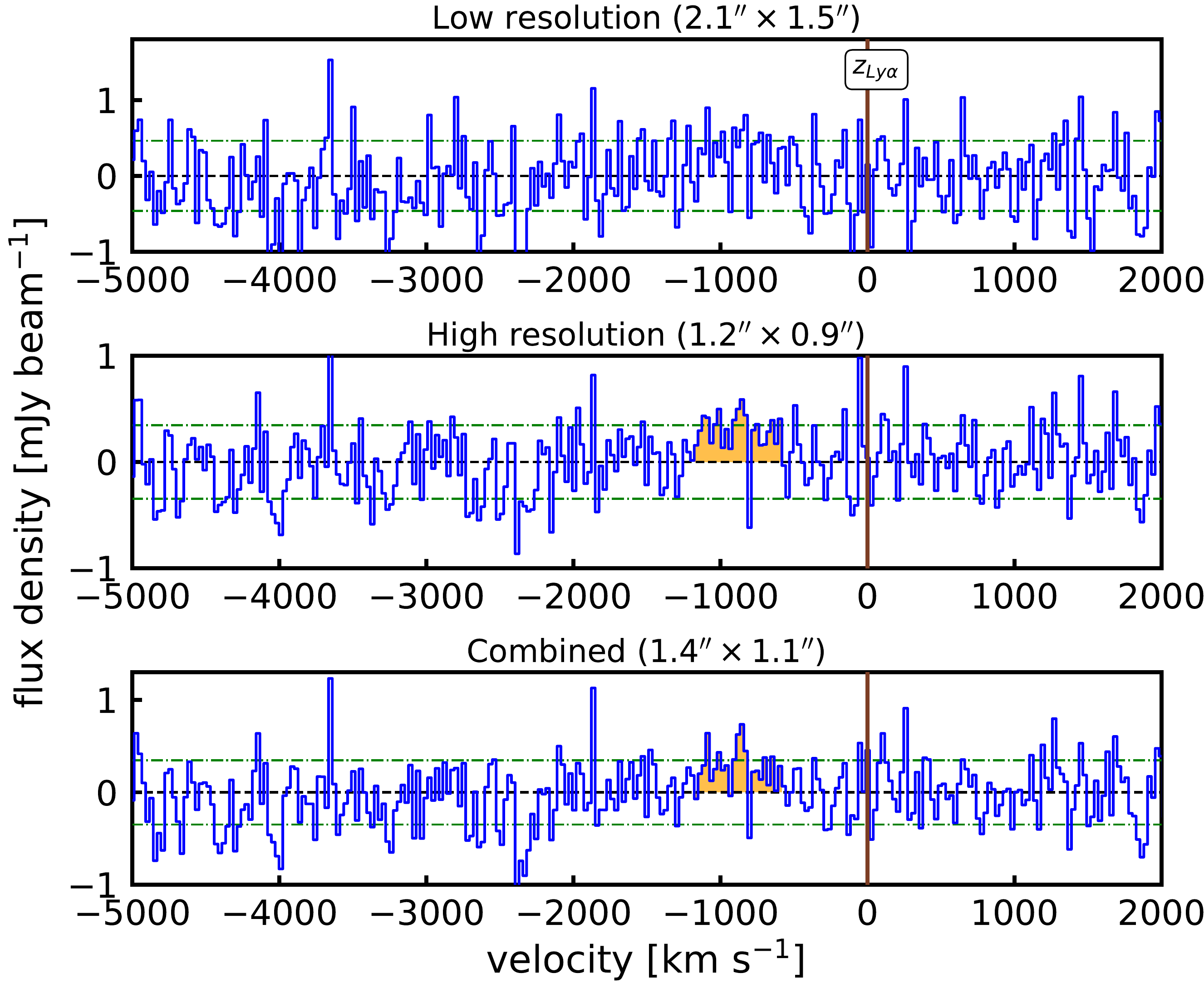}
\end{subfigure}
\caption{\textit{Left:} NOEMA spectrum of GN-108036 with a possible new \cii\ 158 $\mu$m transition detection (orange area). In all three panels, the green dotted line indicates the respective rms noise for three different data sets. Red solid line indicates the redshift measured by Ly$\alpha$ detection. \textit{Right:} Same as the left panel but for the companion system.}
\label{fig:spectras}
\end{figure*}

Fig.~\ref{fig:spectras} shows the \cii\ line spectra of GN-108036 (left) and the potential companion (right) extracted from the D, C, and combined array configuration data, respectively. 

\begin{figure}[htbp!]
    \centering
    \includegraphics[scale=0.3]{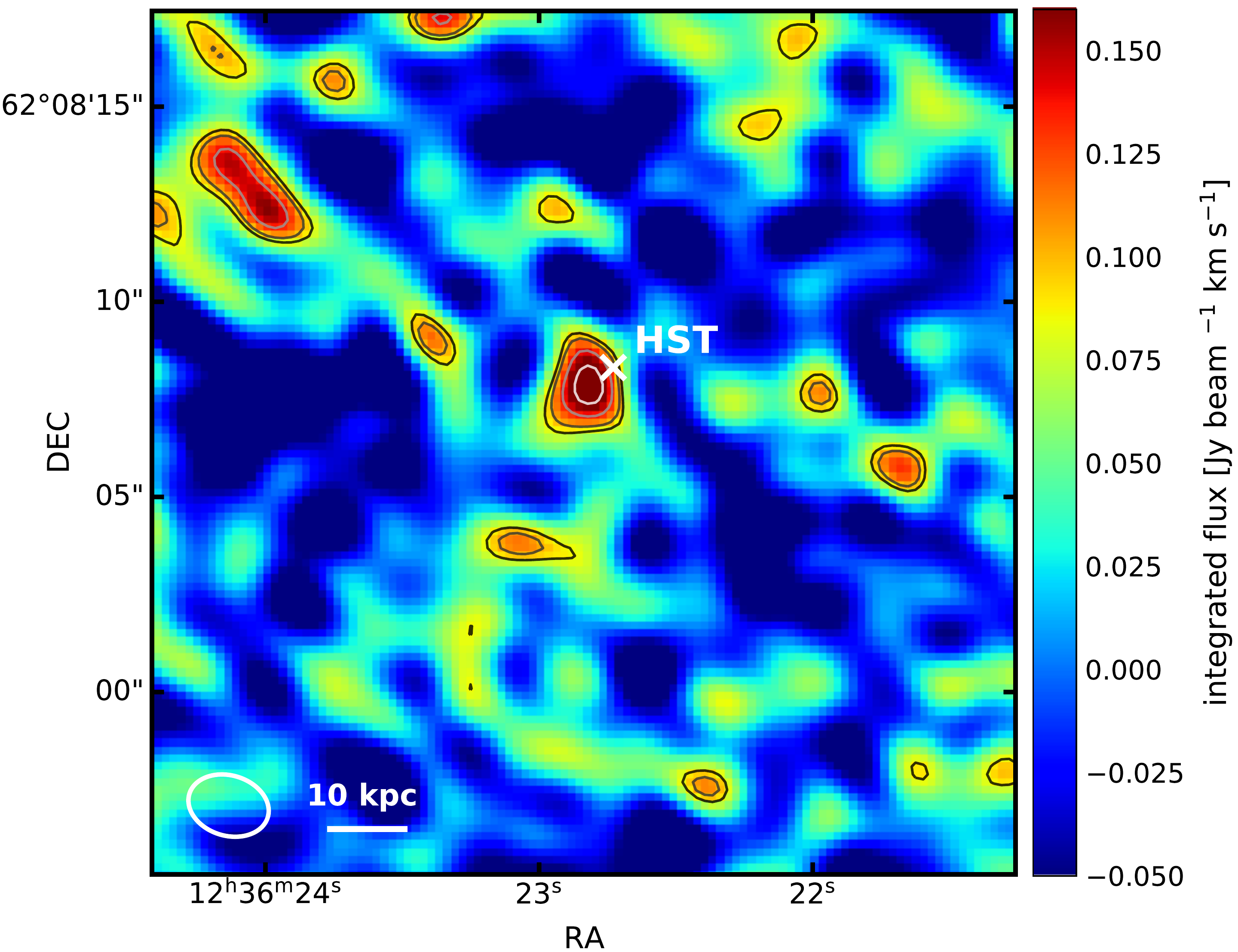}
    \caption{Flux map of GN-108036 in [CII] emission line for the compact data set. The contours corresponds to the 2.5$\sigma$, 3$\sigma$, 3.5$\sigma$, 4$\sigma$, 4.5$\sigma$, 5$\sigma$ and 5.5$\sigma$ (integrated) levels.  The beam size is plotted in the bottom left.}
    \label{fig:flux_maps}
\end{figure}

Fig.~\ref{fig:flux_maps} shows the \cii\ line integrated intensity map of GN-108036 based on the compact array (D) data. The contours correspond to 2.5, 3, 3.5, 4, 4.5, 5 and 5.5$\sigma$ significance levels. The white cross at the center corresponds to the position of the HST rest-frame UV emission from GN-108036.

\end{appendix}

\end{document}